\documentclass[aps,preprint,epsfig,rotate]{revtex4}
\usepackage{graphicx}
\usepackage{bm}
\usepackage{epsfig}

\input epsf
\begin{document}
\title{\bf Thermonuclear burn-up in deuterated methane CD$_4$.}

 \author{Alexei M. Frolov}
 \email[E--mail address: ]{afrolov@uwo.ca}

\affiliation{Department of Chemistry \\
 University of Western Ontario, London, Ontario N6H 5B7, Canada}

\date{\today}

\begin{abstract}

The thermonuclear burn-up of highly compressed deuterated methane CD$_4$ is
considered in the spherical geometry. The minimal required values of the
burn-up parameter $x = \rho_0 \cdot r_f$ are determined for various
temperatures $T$ and densities $\rho_0$. It is shown that thermonuclear
burn-up in CD$_4$ becomes possible in practice if its initial density
$\rho_0$ exceeds $\approx 5 \cdot 10^3$ $g \cdot cm^{-3}$. Burn-up in
CD$_2$T$_2$ methane requires significantly ($\approx$ 100 times) lower
compressions. The developed approach can be used in order to compute the
critical burn-up parameters in an arbitrary deuterium containing fuel.

\end{abstract}

\maketitle

\newpage

In this communication the thermonuclear burn-up in highly compressed
CD$_4$ (and CD$_2$T$_2$) methane is considered. Our goal is to determine the
minimal required values for the burn-up parameter $x$ (where $x = \rho_0
\cdot r_f$), density $\rho_0$ and temperature $T$ at which thermonuclear
burning starts locally and then propagates successfully to the rest of the
highly compressed CD$_4$ methane. Let us assume that initially we have the
infinite, homogeneous, immovable CD$_4$ (and CD$_2$T$_2$) methane (gas),
which is assumed to be cold at time $t = 0$. The burn-up problem is
considered in the following form. Inside homogeneous immovable CD$_4$
methane a quite large amount of energy $Q_0$ is instantaneously released in
a very small (point) volume. The temperature in this volume increases very
rapidly to extremely large values. As a result a high-temperature thermal
wave forms, which begins to propagate from this very hot volume into the
rest of the cold methane. After some time this wave transforms into the
shock wave, which propagates and disappears, in the general case, in the
cold methane.

However, if the three following conditions are obeyed: (1) the initial
density of the highly compressed methane $\rho_0$ is quite large, (2) the
temperature in the central, heated region exceeds the ignition temperature
$T \geq T_c$ and (3) its volume is large enough $V \geq V_c$, then the
thermonuclear burn wave may form and propagate out, igniting the rest of the
cold CD$_4$. In general, the thermonuclear burn wave propagates initially as
a thermal wave, and later as a detonation wave, i.e. the shock wave which
supports itself \cite{Avr} - \cite{Fro1}.

In this sudy we want to determine the critical conditions, i.e. the minimal
possible $\rho_0, T_c$ and $V_c$ values at which the combustion zone still
expands to the infinity in the cold CD$_4$ or CD$_2$T$_2$ methane,
respectively. In the general case, this problem is very complicated, but
here we restricted ourselves to the spherical geometry only. This means that
the initial hot spot has a spherical form, and later, the expanding thermal
or detonation waves also remain spherically symmetric. In this case the
arising gas flow is one-dimensional, and therefore, the appropriate
equations can be also written in the one-dimensional form (see below).
Finally, the initially complicated problem of $V_c$ determination is reduced
to the one dimensional and relatively simple problem: to find a critical
value for some linear $r_c$ parameter, which can be called by the critical
radius of the initial hot spot. The consideration of the cylindrical and
planar cases can be based on analogous assumptions, but our present study is
concentrated mainly on the spherical symmetry, since this case is of
specific interest for various applications in practice.

As is well known (see e.g. \cite{Avr}, \cite{Fro1}) high-temperature burn-up
in thermonuclear fuel is governed by the so-called burn-up equation. Let
$r_f$ be the space radius of the hot (i.e. combustion) zone at the time $t$.
The burn-up parameter $x = \rho_{0} \cdot r_{f}$ (where $\rho_0$ is the
initial density) plays a very important role below. The initial values of
$r_f$ and $x$ are designated below as $r_0 = r_c$ and $x_0 = x_c$. This
corresponds to the physical sense of $r_c$ as the minimal or critical radius
of the hot zone for which the thermonuclear burn wave still propagates to
infinity (in the cold CD$_4$ methane). The temperature $T$ behind the shock
or thermal wave (i.e. in the hot or combustion zone) is assumed to be
significantly higher than in the initially cold CD$_4$ methane, where $T_0
\approx 0$. Furthermore, without loss of generality we may assume that $T$
inside of the hot zone does not depend upon spatial coordinates. Likewise,
the ion and electron temperatures are equal to each other and coincide with
$T$. Such an approximation has a quite good accuracy for burn-up problems,
unless the temperature $T$ exceeds $25$ $keV$ \cite{Avr}. In this
approximation the burn-up equation can be written (for detail see e.g.
\cite{Fro1}) in the form which contains only one variable, namely the
burn-up parameter $x = \rho_{0} \cdot r_{f}$ ($g \cdot cm^{-2}$):
\begin{equation}
\frac{d T}{dx} = - \frac{\nu}{x} \cdot T + \frac{q(x,T)}{{\cal C} \cdot
V_{\max}} \label{eq1}
\end{equation}
where $q(x,T)$ is the so-called normalized energy release function, $\nu =$
3, 2 and 1 for the spherical, cylindrical and plane cases, respectively.
${\cal C}$ is the specific heat per unit mass, which is assumed to be a
constant on the spatial coordinates and time. $V_{\max} = \frac{d
r_{f}}{dt}$ is the maximal velocity of the hot (i.e. combustion) zone
expansion. Since such a zone expands either by the detonation or thermal
waves, the velocity $V_{\max}$ is the larger of the two corresponding
velocities, i.e. $V_{\max} = max \{ (\frac{d r_{f}}{dt})_{d}, (\frac{d
r_{f}}{dt})_{t} \}$. In general, it can be shown that the thermal wave may
propagate the thermonuclear burning only at hardly attainable temperatures
$T \geq$ 23 - 25 $keV$ and only in the dense equimolar $DT-$mixture. In all
other cases, including $DT-$mixtures with low ($\leq 10 \%$) tritium
concentrations, highly compressed deuterium, $DT-$hydrides and deuterides of
light elements, etc, the thermonuclear burning (at $T \leq 25$ $keV$)
propagates by using only the relatively slow, high-temperature detonation
expansion. This means that without loss of generality we can assume in the
last equation $V_{\max} = V_D$, where $V_D$ is the velocity of the
detonation wave.

To determine the velocity $V_D$ we shall apply the strong explosion
approximation, which has a good accuracy for the considered problems (see
e.g. \cite{Avr}, \cite{Fro1}). In this approximation the expression for
$V_D$ takes the from
\begin{equation}
 V_{D} = {a(\gamma) \cdot \sqrt{{\cal C} \cdot T}} = {\sqrt{\frac{3
 \cdot \pi \cdot (\gamma + 1)^{2} \cdot (\gamma - 1)}{8 \cdot (3 \cdot
 \gamma - 1)}} \cdot \sqrt{{\cal C} \cdot T}} = 5.28443640 \cdot 10^{-3}
 \cdot \sqrt{{\cal C} \cdot T}
\end{equation}
where $\gamma = \frac{5}{3}$ and ${\cal C}$ is in $MJ \cdot g^{-1} \cdot
keV^{-1}$ ($1 MJ = 1 \cdot 10^{6} J$), $T$ is in $keV$ and $V_{D}$ is in
$cm \cdot nsec^{-1}$. The specific heat ${\cal C}$ (in $MJ \cdot g^{-1}
\cdot keV^{-1}$) per unit mass is of the form \cite{Fro2}, \cite{Fra}:
\begin{eqnarray}
{\cal C} = \frac{144.7164346}{\overline{A}} \cdot ( 1 + \overline{Z} )
\; \; \; \; \; \; \; MJ \cdot g^{-1} \cdot keV^{-1}
\end{eqnarray}
where the $\overline{A}$ and $\overline{Z}$ values are the mean atomic mass
number and the mean atomic number, respectively. They must be expressed in
units in which the proton charge and proton mass equal 1. Also we assume
that the difference between the proton and neutron masses is negligible and
also the mass defect for all considered nuclei equals zero exactly. Both
these approximations are traditional in thermonuclear problems, and they
have been widely used in earlier works (see e.g. \cite{Fra} or \cite{Lov}).
Therefore, for the considered CD$_4$ methane $\overline{Z} = 2.00$ and
$\overline{A} = 4.00$, while for the CD$_2$T$_2$ methane $\overline{Z} =
2.00$ and $\overline{A} = 4.40$. From here one easily finds: ${\cal C} =
108.537260$ $MJ \cdot g^{-1} \cdot keV^{-1}$, $V_D = 5.50539183 \cdot
10^{-2} \cdot \sqrt{T}$ $cm \cdot nsec^{-1}$ for CD$_4$ methane and ${\cal
C} = 98.6702963$ $MJ \cdot g^{-1} \cdot keV^{-1}$, $V_D = 5.24918515 \cdot
10^{-2} \cdot \sqrt{T}$ $cm \cdot nsec^{-1}$ for CD$_2$T$_2$ methane ($T$ is
in $keV$).

To discuss the energy release function $q(x,T)$ in Eq.(\ref{eq1}) we
introduce the useful pair notation $(\alpha,\beta)$ which designates the
$(d,d)-$pair for CD$_4$ methane and $(d,t)-$pair for CD$_2$T$_2$ methane. By
using this notation the corresponding energy release function $q(x,T) =
q_{\alpha,\beta}(x,T)$ Eq.(\ref{eq1}) can be written in the following
universal form (for more detail see \cite{Fro1}):
\begin{eqnarray}
 q_{\alpha,\beta}(x,T) = \mu^{2}_{\alpha,\beta} \cdot s_{\alpha,\beta}(x,T)
    \cdot ( 1 + B_{\alpha,\beta} \cdot \frac{b_{\alpha,\beta} \cdot x}{1 +
  b_{\alpha,\beta} x} ) - \frac{\kappa_{\alpha,\beta}
    \mu^2_{\alpha,\beta} C_{\alpha,\beta} \sqrt{T}}{1 +
  c_{\alpha,\beta} \sqrt{\rho_{0} \cdot x} T^{-\frac{7}{4}}}
  \label{eq4}
\end{eqnarray}
where $\mu_{\alpha,\beta}$ is the ratio of the $(\alpha,\beta)$ fragment
mass to the total atomic (quasi-molecular) mass and the factor
$\kappa_{\alpha,\beta}$ takes the well known form \cite{Lov}:
\begin{eqnarray}
 \kappa_{\alpha,\beta} = \frac{1}{N^2_{\alpha,\beta}}
 \frac{(\sum_i n_i \cdot Z_i) (\sum_i n_i Z^2_i)}
 {(\sum_{\alpha} n_{\alpha} Z_{\alpha})
 (\sum_{\alpha} n_{\alpha} Z^{2}_{\alpha})} =
 \frac{1}{N^2_{\alpha,\beta}} \frac{(\sum_i n_i \cdot Z_i)}{2}
 \frac{(\sum_i n_i Z^{2}_{i} )}{2}
\end{eqnarray}
where the summation is taken over all ions present. The factor
$\frac{1}{N^2_{\alpha,\beta}} = \frac{1}{2^2} = \frac{1}{4}$ corresponds to
the fact that CD$_4$ contains 4 deuterium atoms or two $d-d$ fragments (i.e.
$N_{dd} = 2$). Analogously, there are two $d-t$ fragments in each molecule
of CD$_2$T$_2$ methane (i.e. $N_{dt} = 2$). For instance, from the last
equations for the CD$_2$T$_2$ methane one easily finds that
$\mu_{\alpha,\beta} = \mu_{d,t} = \frac{2 + 2 + 3 + 3}{12 + 2 + 2 + 3 + 3} =
\frac{10}{22}$ and $\kappa_{\alpha,\beta} = \kappa_{d,t} = \frac{1}{4} \cdot
\frac{6 + 1 + 1 + 1 + 1}{2} \cdot \frac{6^2 + 1^2 + 1^2 + 1^2 + 1^2}{2} =
25$. Finally, $\kappa_{d,t} \cdot \mu^2_{d,t} = 5.165289256$. The use of
these two factors $\mu_{\alpha,\beta}$ and $\kappa_{\alpha,\beta}$ allows us
to apply the same short (i.e. ionic) energy release functions
$s_{\alpha,\beta}(x,T)$ and bremsstrahlung constants the $C_{\alpha,\beta}$
factor as for the appropriate pure deuterium or equimolar DT-mixture. At
least in the first order such an approximation seems to be quite good
\cite{Lov}. In general, the parameters $B_{\alpha,\beta}, b_{\alpha,\beta}$
and $c_{\alpha,\beta}$ depend upon the ionic contents of the considered
mixture. But for our present approximate purposes we shall assume that in
the CD$_4$ and CD$_2$T$_2$ methane they are exactly the same as for pure
deuterium and equimolar DT-mixture, respectively. Their numerical values can
be found in \cite{Avr}, \cite{Fro1}.

The short energy release functions $s_{\alpha,\beta}(x,T)$ in Eq.(\ref{eq4})
correspond to the energy gain produced only by positive high-energy ions
arising in the thermonuclear reactions. They can be expressed also in the
following general form:
\begin{equation}
 s_{\alpha,\beta} = Y \cdot (\overline{\sigma_{\alpha \beta} \cdot
 v_{\alpha \beta}}) \; \; \; \; \; \; \frac{MJ \cdot cm^{3}}{g^2 \cdot nsec}
\end{equation}
where, e.g. $\sigma_{dt}$ is the $(d,t)-$reaction cross-section, $v_{dt}$ is
the relative velocity of the deuterium and tritium nuclei and
$(\overline{\sigma_{dt} \cdot v_{dt}})$ means the averaged (dimensionless)
value over the whole range of relative velocities. The numerical factors $Y$
are \cite{Fro1}: $8.17611 \cdot 10^{18}$ for the $(d,t)-$reaction and
$3.03891 \cdot 10^{19}$ for the $(d,d)-$reaction when all reactions for
${}^{3}He$ nuclei are ignored. However, if the $(d,d)-$reaction is
considered in the presence of high intensity neutron fluxes, then the factor
$Y$ equals $4.63191 \cdot 10^{19}$. The last case represents the situation
when all ${}^{3}$He nuclei react with neutrons, and the following
$(d,t)-$reactions are included in the consideration. Now, the remaining
problem is to find an analytical or numerical expression for the
$(\overline{\sigma_{dd} \cdot v_{dd}})$ and $(\overline{\sigma_{dt} \cdot
v_{dt}})$ values which depend upon the temperature $T$ only. Presently,
they have been chosen from \cite{Fro1}, but it should be mentioned that the
analogous $(\overline{\sigma_{dt} \cdot v_{dt}})$ function from \cite{Avr}
and  $(\overline{\sigma_{dd} \cdot v_{dd}})$ function from \cite{Bru}
produce quite close results.

Now, the solution of the burn-up equation Eq.(\ref{eq1}) not contain any
principle difficulties. Moreover, this solution can be written in the
following integral form \cite{Fro1}:
\begin{equation}
 T(x) = T_0 \cdot ( \frac{x_0}{x} )^{\nu} + \frac{1}{{\cal C} \cdot
 x_0^{\nu}} \cdot \int_{x_0}^{x} dy \cdot \frac{q(y,T(y)) \cdot
 y^{\nu}}{V_D(y,T(y))}
\end{equation}
where $x_0 = \rho_0 \cdot r_0$ and $T(x_0) = T_0$ are the initial value of
the burn-up parameter and temperature, respectively. The general $T(x)$
dependence has the following qualitative behaviour, when $x$ grows: it
starts at $x = x_0$ ($T(x_0) = T_0$), then decreases, reaches a minimum and
later increases to infinity. The minimal temperature $T_{min}$ can not be
negative and the appropriate condition $T_{min} = 0$ is used to determine
the critical value $x_0 = x_c (= \rho_0 \cdot r_c)$ of the burn-up
parameter.

The results of our numerical calculations are presented in Table I (for
CD$_4$ methane) and Table II (for CD$_2$T$_2$ methane). These Tables contain
the $x_c$ values for various temperatures $T$ (4 $keV$ $\leq T \leq$ 20
$keV$) and different initial densities $\rho_0$. In Table I the two subcases
are considered for each temperature and density. The left subcolumn $(a)$ in
Table I represents the results when all reactions with ${}^3He$ nuclei are
ignored. The right subcolumn $(b)$ in this Table corresponds to the case,
when all arising ${}^3He$ nuclei react with neutrons, i.e. the thermonuclear
burn-up in the presence of high intense neutron fluxes. In the last case we
assume that the $(n,{}^{12}$C$;{}^{13}$C$,\gamma)$ reaction has relatively
small cross-sections. As it follows from Table I the propagation of the
thermonuclear detonation wave in CD$_4$ methane simplifies significantly in
the presence of high intense neutron fluxes. This effect is of paramount
importance for relatively small (or realistic) temperatures $T \approx$ 4 -
6 $keV$. In these cases the critical burn-up parameter drops by
approximately 3 times. The thermonuclear detonation in CD$_2$T$_2$ methane
(see Table II) is not very sensitive to the presence (or absence) of any
neutron fluxes.

In discussing the results from Tables I and II, we note that the
thermonuclear burn-up in CD$_4$ methane requires significantly greater
compression than in CD$_2$T$_2$ methane. For realistic temperatures $T
\approx 4 - 6$ $keV$ the deuterated CD$_4$ methane must be compressed
$\approx$ 75 - 100 times more strongly than CD$_2$T$_2$. In other words, the
thermonuclear detonation in CD$_4$ methane may propagate successfully only
if its initial densities $\rho_0 \approx 2.5 \cdot 10^3 - 5 \cdot 10^3$ $g
\cdot cm^{-3}$ or higher, which are quite close to the corresponding Fermi
limits (see below). Analogous 'critical' densities for CD$_2$T$_2$ methane
are $\approx$ 70 - 90 $g \cdot cm^{-3}$ and quite comparable with the
densities required in order to produce thermonuclear burn-up in the dense
deuterium or $DT-$mixtures with very low ($\leq 1 \%$) tritium
concentrations (see \cite{Fro1}). It should be mentioned also that the
minimal energy ${\cal E}_c$, which is needed to produce thermonuclear
ignition in the spherically symmetric case, can be easily evaluated in
terms of the known $x_c$ parameter or critical radius $r_c$: ${\cal E}_c =
\frac{4\pi}{3} \cdot x_c^3 \cdot \rho_0^{-2} \cdot {\cal C} \cdot T =
\frac{4\pi}{3} \cdot r_c^3 \cdot \rho_0 \cdot {\cal C} \cdot T$, where
$\rho_0$ is the initial density and $T$ is the burn-up temperature.

Note also, that the extremely violated energy balance in the deuterated
CD$_4$ methane produces another complication which can not be found in
CD$_2$T$_2$. The source of such a complication is obvious. Indeed, the
velocity of the detonation wave $V_{D}$ increases with the temperature as
$V_{D} \simeq a \cdot \sqrt{T}$, while the thermonuclear ignition in CD$_4$
takes obviously a significantly longer time than in CD$_2$T$_2$. Therefore, we
would expect that at some critical temperature $T_{cr}$ the shock wave
breaks away from the burn wave and propagates into the cold fuel without
producing any ignition. Since the energy losses related with the expansion
of the hot zone are not compensated, the temperature behind the shock wave
decreases rapidly $T \simeq r_f^{-3}$. Later, when the spatial radius of the
hot zone increases to the $r_{cr}$ value and its temperature drops to the $T
= T_{cr}$ value, then the thermonuclear burn-up begins, i.e. the initial,
very fast shock wave slows down and transforms into the detonation wave with
a lower temperature. In CD$_4$ methane such a critical temperature $T_{cr}
\approx 11.5 - 13.75 keV$ and its value depends on the initial density
$\rho_{0}$. The existence of the critical temperature means also that in
CD$_4$ methane the critical burn-up parameter $x_{c}(T,\rho_{0})$ has the
finite, non-zero limit at $T \rightarrow +\infty$: $\lim_{T \rightarrow
+\infty}\overline{x}(T,\rho_{0}) = \overline{x}(T_{cr},\rho_{0})$. In other
words, for CD$_4$ methane the critical burn-up radius $r_{cr} (r_{cr} =
\frac{\overline{x}(T_{cr},\rho_{0})}{\rho_{0}})$ can be determined uniformly
for each density, and moreover, its value: (1) does not vanish when
temperature grows, and (2) does not depend on the temperature $T$, if $T
\geq T_{cr}$. In fact, this means that the non-physical part of the burn-up
curve (where $\frac{d x_c}{d T} > 0$) is ignored and replaced by the
straight line $\frac{d x_c}{d T} = 0$. For the CD$_2$T$_2$ methane $\frac{d
x_c}{d T} < 0$ everywhere (at $T \leq 20 keV$) and the principal difference
between the thermonuclear burn-up in CD$_4$ and CD$_2$T$_2$ can be easily
understood from comparison of Tables I and II. The $\overline{x}$ and
$T_{cr}$ values are also presented (for each density) in Table I. Since
$T_{cr} \leq 13.75 keV$ in all cases, the higher temperatures ($T \geq 14$
$keV$) are not considered in Table I.

Now, it is necessary to show that the obtained solution corresponds to the
actual burn-up process. In general, successful burn-up means that the two
following conditions are obeyed: firstly, the energy release in the hot zone
behind the expanding thermonuclear burn wave must exceed all energy losses
related with the hot zone expansion. This gives the burn-up equation
Eq.(\ref{eq1}) used above. Secondly, the thermonuclear burn wave must move
faster than the following disassembly (or rarefaction) wave. The later is
essentially the high-temperature sound wave. Its velocity $C_s$ takes the
following form:
\begin{equation}
 C_s(T) = \sqrt{\frac{\gamma_i \cdot P_i + \gamma_e \cdot P_e}{\rho_0}} =
 \sqrt{\frac{2}{3} \cdot \gamma \cdot {\cal C} \cdot T} \label{eq8}
\end{equation}
where $T$ is in $keV$, $\gamma = \frac{5}{3}$ and ${\cal C}$ is determined
above. $P_i = n_i \cdot T$ and $P_e = n_e \cdot T$ are the ion and electron
pressures, respectively. As follows from Eq.(\ref{eq8}) in the strong
explosion approximation we always have $V_D(T) > C_s(T)$ for arbitrary $T$.
In other words, in this approximation the disassembly wave always moves
behind the thermonuclear burn wave, i.e. in the combustion zone. The motion
of the disassembly wave in the combustion area does not mean the
termination of thermonuclear burning. However, if the disassembly wave
passes the thermonuclear burn wave, then the fuel will be disassembled
before the thermonuclear burning may start effectively. Such a situation
can be found for both relatively small ($\approx 1 keV$) and very high ($>
100 keV$) temperatures.

In particular, at small temperatures and high compressions the electrons are
very close to the Fermi limit of electron degeneracy. As a result, in this
case, the velocity of the disassembly wave is significantly larger than it
follows from Eq.(\ref{eq8}). Indeed, in this limit the electron pressure is
given by the same formula $P_e = n_e \cdot \overline{T}_e$, where the
effective electron temperature $\overline{T}_e$ takes the form:
\begin{eqnarray}
 \overline{T}_e = T_{ef} \cdot \Bigl[ 1 + \frac{\pi^2}{15} \cdot
 (\frac{T_e}{T_{ef}})^2 + \frac{7 \pi^4}{150} \cdot (\frac{T_e}{T_{ef}})^4
 + \ldots \Bigr]
\end{eqnarray}
where $T_{ef}$ is the equivalent Fermi temperature, given by (see e.g.
\cite{LL}):
\begin{eqnarray}
 T_{ef}(keV) = \frac{2}{5} \cdot ( \frac{3 N_A}{8 \pi} )^{\frac{2}{3}} \cdot
 \frac{h^2}{2 m_e} \cdot ( \frac{\rho_0 \cdot \overline{Z}}{\overline{A}}
 )^{\frac{2}{3}} = 9.3019938991 \cdot 10^{-3} \cdot
 (\frac{\overline{Z}}{\overline{A}})^{\frac{2}{3}} \cdot
 (\rho_0)^{\frac{2}{3}}
\end{eqnarray}
where $N_A, h$ and $m_e$ are the Avogadro number, Planck constant and
electron mass, respectively \cite{COD}. For instance, for ${}^{12}CD_4$
methane with $\rho_0 = 1 \cdot 10^6$ $g \cdot cm^{-3}$ one easily finds
$T_{ef} = 58.50$ $keV$. Therefore, for $T_e = 1$ $keV$ the effective
electron temperature $\overline{T}_e \approx T_{ef}$ and actual electron
pressure $P_e$ exceeds its classical value in 58.60 times. In this case,
the disassembly wave moves in $\approx 5$ times faster than it follows
from Eq.(\ref{eq8}). Briefly, we can say that the appropriate correction on
the electronic degeneracy is obviously needed when the actual temperature $T
(= T_e)$ is comparable with the equivalent Fermi temperature $T_{ef}$
\cite{LL}. In our present analysis the largest $T_{ef}$ value can be found
in CD$_4$ methane, where $\rho_0 \le 1 \cdot 10^4$ $g \cdot cm^{-3}$, i.e.
$T_{ef} \leq 0.27199195$ $keV$. This indicates that the maximal $T_{ef}$
temperature is significantly smaller than temperatures used in the present
study $(T \ge 4$ $keV$), i.e. the appropriate Fermi correction to the
electron pressure $P_e$ is very small and can be ignored.

In conclusion, it should be mentioned that the presented approach predicts
that the thermonuclear burn-up is quite likely also in the higher deuterated
alkanes (C$_n$D$_{2n+2}$), alkenes (C$_n$D$_{2n}$), alkynes
(C$_n$D$_{2n-2}$), etc. Indeed, the both factors $\kappa_{d,d}$ and
$\mu_{d,d}$ have the finite limits when $n \rightarrow \infty$ (these limits
equal $\frac{1}{4}$ and $\frac{25}{16}$, respectively). The determination of
the critical parameters for each of these polymers is straightforward.
Moreover, the critical burn-up parameters for an arbitrary deuterium
containing fuel can be computed by using the formulas presented above. In
particular, for all deuterides of light elements with $Z \leq 9$
(D${}^{19}F$ is included) such an analysis can be found in our paper
\cite{FF}.

\begin{table}[tbp]
\caption{The critical values of the burn-up parameter $x_c$ (in $g \cdot
         cm^{-2})$ for various densities $\rho_0$ (in $g \cdot cm^{-3})$,
         temperatures $T$ (in $keV$) for CD$_4$ methane. The superscripts
         mean that: $(a)$ all reactions with the ${}^{3}$He nuclei are
         ignored, and $(b)$ the thermonuclear burn-up proceeds in very
         intense neutron fluxes.}
  \begin{center}
     \scalebox{0.95}{%
     \begin{tabular}{ccccccc}
      \hline\hline
     & $\rho_0^{(a)}$ & $\rho_0^{(b)}$ & $\rho_0^{(a)}$ & $\rho_0^{(b)}$ &
      $\rho_0^{(a)}$ & $\rho_0^{(b)}$ \\
         \hline
 $T$ & $1 \cdot 10^3$ & $1 \cdot 10^3$ & $5 \cdot 10^3$ & $5 \cdot 10^3$ &
      $1 \cdot 10^4$ & $1 \cdot 10^4$ \\
        \hline
 4  & 13125.467 & 4330.014 & 2834.009 & 986.851 & 1542.655 & 564.471 \\
 5  &  9340.429 & 3074.378 & 2019.777 & 703.628 & 1100.985 & 403.783 \\
 6  &  7783.949 & 2552.553 & 1679.783 & 583.248 &  913.755 & 334.134 \\
 7  &  7010.860 & 2288.540 & 1508.037 & 520.944 &  817.637 & 297.394 \\
 8  &  6585.202 & 2138.550 & 1411.503 & 484.666 &  762.567 & 275.570 \\
 9  &  6341.346 & 2047.877 & 1354.585 & 462.082 &  729.256 & 261.671 \\
 10 &  6206.550 & 1992.443 & 1321.531 & 447.699 &  709.104 & 252.551 \\
 11 &  6145.359 & 1960.466 & 1304.593 & 438.793 &  697.844 & 246.629 \\
 12 &  6135.744 & 1945.657 & 1299.618 & 433.884 &  693.165 & 243.037 \\
 13 &  6135.744 & 1943.435 & 1299.618 & 432.106 &  692.812 & 241.259 \\
       \hline
 $\overline{x}$ & 6135.744 & 1943.435 & 1299.618 & 432.073 & 692.812 &
                   240.914 \\
       \hline
 $T_{cr}$ & 11.62 & 12.60 & 11.99 & 13.16 & 12.38 & 13.71 \\
      \hline\hline
  \end{tabular}}
  \end{center}
  \end{table}
%
%
  \begin{table}[tbp]
    \caption{The critical values of the burn-up parameter $x_c$ (in $g
             \cdot cm^{-2})$ for various densities $\rho_0$ (in $g
             \cdot cm^{-3})$, temperatures $T$ (in $keV$) for CD$_2$T$_2$
             methane.}
     \begin{center}
     \scalebox{0.95}{%
     \begin{tabular}{llllllll}
      \hline\hline
    & $\rho_0$ & $\rho_0$ & $\rho_0$ &
    & $\rho_0$ & $\rho_0$ & $\rho_0$ \\
        \hline
$T$ & $50$ & $100$ & $200$ & $T$ & $50$ & $100$ & $200$ \\
        \hline
 4  & 96.6663 & 60.2729 & 39.7280 & 10 & 8.4305 & 7.2564 & 6.1622 \\
 5  & 50.6829 & 34.2076 & 24.0097 & 11 & 6.6828 & 5.9122 & 5.1495 \\
 6  & 31.1823 & 22.4731 & 16.5634 & 13 & 4.5962 & 4.2287 & 3.8281 \\
 7  & 20.9295 & 15.9492 & 12.2564 & 15 & 3.5037 & 3.3001 & 3.0624 \\
 8  & 14.8525 & 11.8714 &  9.4686 & 18 & 2.6749 & 2.5653 & 2.4291 \\
 9  & 10.9912 &  9.1474 &  7.5431 & 20 & 2.3703 & 2.2870 & 2.1811 \\
      \hline\hline
  \end{tabular}}
  \end{center}
  \end{table}

\begin{thebibliography}{10}

\bibitem{Avr} E.V. Avrorin, L.P. Feoktistov and L.I. Shibarshov, Fiz.
Plasmy. \textbf{6}, 965 (1980) [Sov. J. Plasma Phys., \textbf{6}, 527
(1980)].

\bibitem{Fro1} A.M. Frolov, Plasma Phys. and Contr. Fusion. \textbf{40},
1417 (1998).

\bibitem{Fro2} A.M. Frolov, Can. J. Phys. \textbf{26}, 33 (1994).

\bibitem{Kor} V.P. Korobeinikov, \textit{Problems of point-blast theory},
(American Institute of Physics, New York, 1991).

\bibitem{Fra} G.S. Fraley, E.L. Linnebur, R.K. Mason and R.L. Morse, Phys.
of Fluids. \textbf{17}, 474 (1974).

\bibitem{Lov} S. Glasstone and R.H. Lovberg, \textit{Controlled
Thermonuclear Reactions}, (Van Nostrand, Princeton, New Jersey, 1960).

\bibitem{Bru} K.A. Brueckner and S. Jorna, Rev. Mod. Phys. \textbf{46}, 325
(1974).

\bibitem{LL} L.D. Landau and E.M. Lifshitz, \textit{Statistical Physics},
(Pergamon Press LTD., London, 1958), Chps. V and XI.

\bibitem{COD} E.R. Cohen and B.N. Taylor, Physics Today. \textbf{45} (8), 9
(1993).

\bibitem{FF}A.M. Frolov, V.H. Smith, Jr. and G.T. Smith, Can. J. Phys.
\textbf{80}, 43 (2002).
\end{thebibliography}
\end{document}